\documentstyle[preprint,aps]{revtex}
\begin {document}
\draft
\preprint{UCI-TR 94-30}
\title{Baryons With Two Heavy Quarks as Solitons}
\author{Myron Bander\footnote{Electronic address: mbander@funth.ps.uci.edu}
and
Anand Subbaraman\footnote{Electronic address: anand@funth.ps.uci.edu}}
\address{
Department of Physics, University of California, Irvine, California
92717}

\date{July\ \ \ 1994}
\maketitle
\begin{abstract}
Using the chiral soliton model and heavy quark symmetry we study baryons
containing two heavy quarks.  If there exists a stable (under strong
interactions) meson consisting of two heavy quarks and two light ones, then
we find that there always exists a state of this meson bound to a chiral
soliton and to a chiral anti-soliton, corresponding to a two heavy quark
baryon and a baryon containing two heavy anti-quarks and five light quarks,
or a ``heptaquark".
\end{abstract}


Recently a picture has emerged in which baryons containing a heavy
quark, $Q$, and two light quarks, $q$'s, are viewed as bound states of
a heavy meson made out of $Q{\bar q}$ in the field of a chiral soliton
\cite{CK,Cal,Syr,Kor}. The heavy mesons are discussed using the Isgur-Wise
heavy quark symmetry \cite{IW}; in this formalism all particles with
spins made by combining a fixed spin for the light and for the heavy
quarks are described by a single heavy field creating or annihilating
particles of fixed four velocity. These ideas have been extended to
bound states of a heavy anti-meson, ${\bar Q}q$ in the field of a
chiral soliton \cite{Oh}; this would lead to baryons (not anti-baryons)
with charm $C=-1$ or beauty $B=+1$. In a conventional quark picture
such states are made out of a heavy anti-quark and four light quarks
and have previously been discussed \cite{pentaquark} and named
``pentaquarks''.

In this work we extend these ideas to baryons containing two heavy
quarks. There are interesting differences from baryons with one heavy quark:
in the latter case the ground state is, in the language of
Ref.~\cite{CK}, in the $l=1$ channel. For baryons with two heavy
quarks the ground state consists of a superposition of $l=0$ and $l=2$
states (and possibly other states); independent of the sign of the
potential, at least one state is bound for both the soliton and
anti-soliton. The latter correspond in the conventional picture to
states of two heavy anti-quarks and five light quarks or
``heptaquarks''.

A crucial ingredient for the validity of this picture is the existence of
stable, under strong interactions, bound states of
$Q{\bar q}-Q{\bar q}$ mesons. A
conservative estimate, using the one pion exchange potential between
these particles, gives a binding for the $B^{*}B$ system
\cite{MW}. Bound states for the $DD$ and $BD$ systems are not
excluded.
It is likely that for sufficiently large $m_Q/m_q$ these
states will be bound in a configuration where the heavy and light
systems are in antisymmetric color combinations \cite{Richard}. For
the case where there are stable $QQ{\bar q}{\bar q}$ systems we expect
$QQq$ baryons with spin-parity $\frac{1}{2}^+$; our results are
consistent with this expectation and we lokewise predict the existence
of heptaquarks consisting of two ${\bar Q}$'s and five light quarks.

We analyze the $QQ{\bar q}{\bar q}$ system by first coupling the heavy
quarks to each other, the light anti-quarks to each other and then
combining the two. As mentioned earlier the favored color
configuration is for the $QQ$ system to be in a color ${\bar{\mbox{\bf 3}}}$
and the ${\bar q}{\bar q}$ in a color {\bf{3}}. We shall also consider the
possibility that they are in a symmetric color combination, namely {\bf 6}
and ${\bar{\mbox{\bf 6}}}$. First we shall look at the case where the two
heavy quarks are identical. Then for the antisymmetric color combination the
heavy quark spin $S_H=1$ and we have the following possibilities:
\begin{equation}
\begin{array}{lcccc}
(a)\ \ \ &S_l=1;&I=1;&J=0,1,2;\ \ \ &T_{iJ}^{\alpha}\\
(b)\ \ \ &S_l=0;&I=0;&J=1;\ \ \ &V_J
\end{array}
\label{antisymm}
\end{equation}
For the color symmetric case, $S_H=0$ and:
\begin{equation}
\begin{array}{lcccc}
(a)\ \ \ &S_l=0;&I=1;&J=0;\ \ \ &S^{\alpha}\\
(b)\ \ \ &S_l=1;&I=0;&J=1;\ \ \ &{V_i}
\end{array}
\label{symm}
\end{equation}
In the above $S_l$ is the spin and $I$ is the isospin of the light
anti-quarks and $J$ is the total spin of the $QQ{\bar q}{\bar q}$ system. In
the last column of the above equations we indicate the notation for
heavy fields with three velocity zero that combine fields with fixed
$QQ$ and ${\bar q}{\bar q}$ spin configurations. The lower case indices refer
to light spin  degrees of freedom and the upper case ones to the spin of the
heavy quark combinations; isospin is indicated by upper Greek indices. The
tensor fields $T$ combine a spin zero, $s$, spin one, $v_i$
and symmetric traceless spin two, $t_{ij}$, fields into a
spin multiplet ``superfield''
\begin{equation}
T_{iJ}=\frac{s}{{\sqrt 3}}\delta_{iJ}+\frac{\epsilon_{iJk}v_k}{{\sqrt 2}}
+t_{iJ}\, .\label{heavyfield}
\end{equation}
Heavy quark spin symmetry demands invariance under rotations of the
upper case spin indices.

With $\Sigma$, a unitary $2\times 2$ matrix describing the light Goldstone
pions and $\xi=\Sigma^{\frac{1}{2}}$, the Lagrangian for the heavy
system (with zero three velocity) is
\begin{eqnarray}
{\cal L}=&-&iT_{iJ}^{\alpha\dag}D_t^{\alpha\beta}T_{iJ}^{\beta}-
iV_J^{\dag}\partial_tV_J
 -iS^{\alpha\dag}D_t^{\alpha\beta}S^{\beta}-
iV_i^{\dag}\partial_tV_i\nonumber\\
   &+&g_1\epsilon^{\alpha\beta\gamma}\epsilon_{ijk}T_{iJ}^{\alpha\dag}
     A_j^{\beta}T_{kJ}^{\gamma}+g_2(V_J^{\dag}A_i^{\alpha}T_{iJ}^{\alpha}
      +\mbox{\rm h.c.})+g_3(S^{\alpha\dag}A_i^{\alpha}V_i+\mbox{\rm
        h.c.}) \, .
	  \label{Lagr}
\end{eqnarray}
In the above
\begin{eqnarray}
D_t^{\alpha\beta}&=&\delta^{\alpha\beta}\partial_t-\frac{1}{2}\epsilon^
{\alpha\beta\gamma}
\mbox{\rm tr}\left [\tau^{\gamma}
(\xi^{\dag}\partial_t\xi+\xi\partial_t\xi^{\dag})\right ]
\nonumber\\
A_i^{\alpha}&=&\frac{i}{2}\mbox{\rm tr}\left [\tau^{\alpha}
(\xi^{\dag}\partial_i\xi-\xi\partial_i\xi^{\dag})\right ]\, .
\end{eqnarray}
For completeness we also write down the version of
the Lagrangian for arbitrary four-velocity $\mbox{\rm v}^\mu$ of the
heavy degrees of freedom as
\begin{eqnarray}
{\cal L}_{\mbox{\rm v}}=&-&iT^{\alpha\dag\,\mu\nu} \mbox{\rm v}\cdot
D^{\alpha\beta}
T_{\mu\nu}^{\beta}-
iV^{\dag\,\mu} \mbox{\rm v} \cdot \partial\, V_\mu
 -iS^{\alpha\dag} \mbox{\rm v} \cdot D^{\alpha\beta} S^{\beta}
-iV^{\dag\,\mu} \mbox{\rm v} \cdot \partial \, V_\mu\nonumber\\
   &+&g_1\epsilon^{\alpha\beta\gamma}\epsilon^{\mu\nu\rho\sigma}
T_{\mu\delta}^{\alpha\dag}
     A_\nu^{\beta}T_{\rho\delta}^{\gamma} \,\mbox{\rm v}_\sigma+
g_2(V^{\dag\,\mu}A^{\alpha \nu}T_{\nu\mu}^{\alpha}
      +\mbox{\rm h.c.})+g_3(S^{\alpha\dag}A^{\alpha \mu}V_\mu+\mbox{\rm
        h.c.}) \, .
	  \label{4dLagr}
\end{eqnarray}
In the above  $\alpha,\beta,\gamma$ are isospin indices
and the field $T$ is
\begin{eqnarray}
T_{\mu\nu}&=&\frac{s}{{\sqrt 3}}(-g_{\mu\nu}+\mbox{\rm v}_\mu
\mbox{\rm v}_\nu) +
\frac{\epsilon_{\mu\nu\rho\sigma}v_\rho \mbox{\rm v}_\sigma}{{\sqrt 2}}
\nonumber \\
{}&+&t_{\mu\nu}- t_{\mu\rho}\mbox{\rm v}^\rho \mbox{\rm v}_\nu
- t_{\rho\nu}\mbox{\rm v}^\rho \mbox{\rm v}_\mu
+ t_{\rho\sigma}\mbox{\rm v}^\rho \mbox{\rm v}^\sigma \mbox{\rm v}_\mu
\mbox{\rm v}_\nu \,
,\label{4dheavyfield}
\end{eqnarray}
and satisfies the constraints $\mbox{\rm v}^\mu T_{\mu\nu}=
\mbox{\rm v}^\nu T_{\mu\nu}= 0$.

For the classical $SU(2)\times SU(2)$ soliton \cite{Adkins}
$\Sigma(\vec{r})=\exp\left [i\vec{\tau}\cdot\hat{r}F(r)\right ]$, with
$F(0)=-\pi$, we obtain
\begin{eqnarray}
A_i^{\alpha}(r)&=&a_1(r)\delta_i^{\alpha}+a_2(r)\hat{r}_i\hat{r}^{\alpha}\,
,
\nonumber\\
a_1(r)=\frac{\sin F(r)}{2r}\, ,\ &{}&\ a_2(r)=\frac{rF'(r)-\sin
F(r)}{2r}\, .\label{defA}
\end{eqnarray}

The intrinsic parity of the $QQ{\bar q}{\bar q}$ system is positive
and as we expect the ground state of the baryon to be $\frac{1}{2}^+$
the $QQ{\bar q}{\bar q}$ must be in an even angular momentum state.
For $T_{iJ}^{\alpha}$ and for $V_J$, as defined in
Eq.~(\ref{antisymm}), we can do this with $l=0$ and $l=2$ waves.
\begin{eqnarray}
T_{iJ}^{\alpha}&=&(h_1(r)\delta_i^{\alpha}+h_2(r)\hat{r}_i
\hat{r}_{\alpha})\chi_J \nonumber\\
V_J&=&h_3(r)\chi_J\, ;\label{wavefn}
\end{eqnarray}
$\chi_J$ is the wavefunction for the heavy spin, which decouples from
the problem, as demanded by heavy spin symmetry.
 For the symmetric color configurations,
Eq.~(\ref{symm}), there are {\em no} even parity configurations
possible. It is heartening that it is the color antisymmetric configurations
of $QQ{\bar q}{\bar q}$ that are expected to be bound \cite{Richard}.
In the $h_1,\, h_2,\, h_3$ basis the potential energy matrix is
\begin{equation}
V=2g_1\left( \begin{array}{cccc}
2(3a_1+a_2) &{}& 2a_1 & \frac{g_2}{g_1}(3a_1+a_2)\\
2a_1 &{}& 0 & \frac{g_2}{g_1}(a_1+a_2)\\
\frac{g_2}{g_1}(3a_1+a_2)&{}& \frac{g_2}{g_1}(a_1+a_2)& 0
\end{array} \right )\, .\label{potential}
\end{equation}

Let us first look for the eigenvalues of $V$ for the case $g_2=0$. Two
eigenvalues are of opposite sign and one is at zero; thus one of the
configurations is bound. For small $g_2/g_1$, independent of the sign of
$g_2$ the level at zero repels the other two and one becomes more bound and
the other more unbound. As can be seen in Fig.~\ref{fig1} and
Fig.~\ref{fig2}  this behavior persists at higher values of $g_2/g_1$. We
also note that the minimum eigenvalues occur at $r=0$. We find that
independent of the signs of the couplings we always have at least one bound
state.

Now if we couple the $QQ{\bar q}{\bar q}$ system to the
field of an anti-soliton, $A_i^{\alpha}$ changes sign and the positive
eigenvalues of the soliton case become negative, resulting in
heptaquark states. {\it This analysis predicts that a bound $QQ{\bar
q}{\bar q}$ meson will yield at least one bound $QQq$ and one ${\bar
Q}{\bar Q}qqqqq$ baryon}. Zero mode quantization shows that it is the
soliton that will determine the spin and isospin quantum numbers of these
objects.

We briefly turn to the case when the two heavy quarks are not identical.
The only change is that in both the color symmetric and color antisymmetric
case the heavy quark spin can be both zero and one; there is no effect on
the light quarks and thus the spectrum is doubled. This is consistent with
what we would expect from the usual quark model description of these
baryons.

This research was supported in part by the National Science Foundation under
Grant PHY-9208386.

\nobreak

\begin{figure}
\caption{Eigenvalues of the potential matrix for different values of
$g_2/g_1$. Solid line corresponds to $g_2/g_1=0$ and the dashed one to
$g_2/g_1=1$.}
\label{fig1}
\end{figure}
\begin{figure}
\caption{Same as Fig. 1 for $g_2/g_1=5$, solid line and $g_2/g_1=10$,
dashed line.}\label{fig2}
\end{figure}

\begin{references}
\bibitem{CK}
C.\ Callan and I.\ Klebanov, Nucl.\ Phys. {\bf B262}, 365 (1985).
\bibitem{Cal}
M.\ Rho, D.\ O.\ Riska and N.\ Scoccola, Z.\ Phys.\ A{\bf 341}, 341 (1992);
E.\ Jenkins and A.\ V.\ Manohar, Phys.\ Lett.\ {\bf B294}, 173 (1992);
Z.\ Guralnik, M.\ Luke and A.\ V.\ Manohar, Nucl.\ Phys. {\bf B390},
474 (1993);
E.\ Jenkins, A.\ V.\ Manohar and M.\  Wise, Nucl.\ Phys. {\bf B396},
27,38 (1993).
\bibitem{Syr}
K.\ S.\ Gupta, M.\ A.\ Momen, J.\ Schechter and A.\ Subbaraman, Phys\.
Rev.\ D{\bf 47}, R4835 (1993); A. Momen, J. Schechter and A.
Subbaraman, Syracuse University
Report SU-4240-564, University of California Irvine Report UCI-93-47, to
appear in Phys. Rev.~D.
\bibitem{Kor}
D.\ P.\ Min, Y.\ Oh, B.\ Y.\ Park and M.\ Rho, Seoul Report No.\ SNUTP 92-78;
Y.\ Oh, D.\ Y. Park and D.\ P. Min, Phys.\ Rev.\ D{\bf 49}, 4649 (1994);
M.\ A.\ Nowak, M.\ Rho and I.\ Zahed, Phys. Lett {\bf B303}, 130 (1993);
H.\ K.\ Lee, M.\ A.\ Nowak, M.\ Rho and I.\ Zahed, Ann.\ Phys. {\bf
227}, 175 (1993).
\bibitem{IW}
N.\ Isgur and M.\ B.\ Wise, Phys. Lett. {\bf B232}, 113 (1989); {\it
ibid.\/} {\bf 237}, 507 (1990); others.
\bibitem{Oh}
Y.\ Oh, B.-Y.\ Park and D.-P.\ Min, Seoul National University
preprint, SNUTP-94/06 (unpublished).
\bibitem{pentaquark}
H.\ J.\ Lipkin, Phys.\ Lett.\ {\bf B195}, 484 (1987); C.\ Gignoux, B.\
Silvestre-Brac and  J.\ M.\ Richard,  Phys.\ Lett. {\bf B193}, 323 (1987).
\bibitem{MW}
A.\ V.\ Manohar and M.\ B.\ Wise, Nucl. Phys. B{\bf 399}, 17 (1993).
\bibitem{Tornqvist}
N.\ A.\ T\"{o}rnqvist, Phys. rev. Lett. {\bf 67}, 556 (1991); Z.\
Phys. {\bf C61}, 525 (1994); A.\ V.\ Manohar and M.\ B.\ Wise, Nucl.
Phys. B{\bf 399}, 17 (1993)
\bibitem{Richard}
J.-M.\ Richard, Institut des Sciences Nucl\'{e}aires, Grenoble
preprint (unpublished).
\bibitem{Adkins}
G.\ S.\ Adkins, C.\ R.\ Nappi and E.\ Witten, Nucl. Phys. {\bf B228},
552 (1983).
\end{references}
\end{document}